\documentclass[a4paper,fleqn,usenatbib]{mnras}

\usepackage{subfigure}
\usepackage{rotating}
\usepackage{newtxtext,newtxmath}
\usepackage[T1]{fontenc}
\usepackage{ae,aecompl}
\usepackage{tikz}

\usepackage{graphicx}	
\usepackage{amsmath}	
\usepackage{amssymb}	

\newcommand{\Ms}{M$_{\odot}$}
\newcommand{\Zs}{Z$_{\odot}$}

\newcommand{\Ni}{{\ensuremath{^{56}\mathrm{Ni}}}}

\newcommand{\popIIccilow}{{0}}
\newcommand{\popIIccihigh}{{43}}
\newcommand{\popIIpii}{{5}}

\newcommand{\popIIcczlow}{{1}}
\newcommand{\popIIcczhigh}{{110}}
\newcommand{\popIIpiz}{{10}}
\newcommand{\popIInz}{{1}}
\newcommand{\popIIppiz}{{0}}
\newcommand{\popIItotalz}{{120}}
\newcommand{\popIIIpiz}{{0}}

\newcommand{\popIIccylow}{{0}}
\newcommand{\popIIccyhigh}{{17}}
\newcommand{\popIIpiy}{{2}}

\newcommand{\wfirst}{{\em WFIRST}}
\newcommand{\euclid}{{\em Euclid}}

\title[Lensed SNe in LSST]{Detecting strongly lensed supernovae at $z \sim 5 - 7$ with LSST}

\author[Rydberg et al.]{Claes-Erik Rydberg$^{1}$\thanks{E-mail: mail@utte.nu},
Daniel J. Whalen$^{2}$,
Matteo Maturi$^{1}$,
Thomas Collett$^{2}$,
\newauthor
Mauricio Carrasco$^{1}$,
Mattis Magg$^{1}$,
Ralf S. Klessen$^{1, 3}$
\\
$^{1}$Universit$\ddot{\mbox{a}}$t Heidelberg, Zentrum f$\ddot{\mbox{u}}$r Astronomie, Institut f$\ddot{\mbox{u}}$r Theoretische Astrophysik, Albert-Ueberly-Str. 2, 69120 \\ 
$^{ }$ Heidelberg, Germany \\
$^{2}$Institute of Cosmology and Gravitation, University of Portsmouth, Dennis Sciama Building, Portsmouth PO1 3FX, UK \\
$^{3}$Universit$\ddot{\mbox{a}}$t Heidelberg, Interdisziplin$\ddot{\mbox{a}}$res Zentrum f$\ddot{\mbox{u}}$r Wissenschaftliches Rechnen (IWR), 69120 Heidelberg, Germany \\
}

\date{Accepted XXX. Received YYY; in original form ZZZ}

\pubyear{2018}

\begin{document}
\label{firstpage}
\pagerange{\pageref{firstpage}--\pageref{lastpage}}
\maketitle


\begin{abstract}

Supernovae (SNe) could be powerful probes of the properties of stars and galaxies at high redshifts in future surveys. Wide fields and longer exposure times are required to offset diminishing star formation rates and lower fluxes to detect useful numbers of events at high redshift. In principle, the Large Synoptic Survey Telescope (LSST) could discover large numbers of early SNe because of its wide fields but only at lower redshifts because of its AB mag limit of $\sim$ 24. But gravitational lensing by galaxy clusters and massive galaxies could boost flux from ancient SNe and allow LSST to detect them at earlier times.  Here, we calculate detection rates for lensed SNe at $z \sim$ 5 - 7 for LSST. We find that the LSST Wide Deep Fast survey could detect up to \popIItotalz{} lensed Population (Pop) I and II SNe but no lensed Pop III SNe. Deep-drilling programs in a single 10 deg$^2$ FoV could detect Pop I and II core-collapse SNe and Pop III pair-instability SNe at AB magnitudes of 27 - 28 and 26, respectively. An alternative deep survey over 80 nights with a one-year cadence could find $\sim$ 8 Pop III SNe. 

\end{abstract}




\section{Introduction}
\label{sec:introduction}

High redshift SNe are powerful probes of the properties of stars and their formation rates at early times because they can be observed at great distances and, to some degree, the masses of their progenitors can be inferred from their light curves.  Early SNe could complement gamma-ray bursts (GRBs) as tracers of stellar populations in the first galaxies (\citealt{re12}; see also \citealt{jlj09,get10,pmb11,jeon11,wise12,pmb12,mura13,xu13,ren15}).  They can also constrain scenarios for early cosmological reionization \citep[e.g.,][]{wan04,ket04,oet05,su06,abs06,awb07,wet08b,suh09,wet10} and early chemical enrichment \citep{mbh03,byh03,ky05,ss07,get07,wet08a,bsmith09,jet09b,get10,ss13,brit15,rit16,slud16,chen17a,chen17b,hart18b}.  

In principle, detections of SNe at $z \gtrsim$ 20 could even reveal the properties and formation rates of Population III (Pop III) stars at cosmic dawn \citep{moriya10,tomin11,tet12,hum12,tet13,ds13,ds14} and shed light on the origins of supermassive black holes \citep[SMBHs;][]{awa09,wf12,jlj12a,jet13,hos13,jet14,um16,tyr17,hle17,smidt18,hle18}.  Primordial SNe will impose one of the few direct constraints on the Pop~III initial mass function (IMF) in the near term because the stars themselves will not be visible to the {\em James Webb Space Telescope} \citep[{\em JWST};][]{jwst,jwst2} or to 30 - 40 m telescopes on the ground and because numerical simulations cannot predict their properties from first principles \citep[e.g.,][]{fsg09,fg11,dw12,glov12} \citep[but see][]{rz13,wind18}.

Longer exposure times and larger survey areas are required to detect transients at high redshifts because of their lower luminosities and smaller star formation rates (SFRs) at early epochs. {\em JWST} and the ELTs will have the sensitivity required to detect even the earliest explosions but only in narrow fields of view that may not capture many events (\citealt{hart18a} predict about 1 SN per year). On the other hand, LSST \citep{LSST}, \euclid{} \citep{euclid} and the Wide-Field Infrared Survey Telescope \citep[\wfirst{};][]{wfirst} could harvest far greater numbers of events because of their large survey areas but only at lower redshifts because of their lower sensitivities. Furthermore, because LSST is limited to the $Y$ band in wavelength it cannot detect transients above $z \sim$ 6 - 7, and because \euclid{} and \wfirst{} are limited to the $H$ band they cannot detect SNe at $z \gtrsim$ 13 - 14. This is due to the fact that flux blueward of the Lyman-$\alpha$ wavelength in the source frame of an event at $z >$ 6 is absorbed by the neutral intergalactic medium (IGM) prior to the end of cosmological reionization. The optimum bands for detecting SNe at $z \sim$ 20 are in the near infrared (NIR) at 2 - 5 $\mu$m \citep{wet12a,wet12c,wet12b}.

But gravitational lensing by galaxy clusters and massive galaxies could offset the lower sensitivities of wide-area surveys and reveal more transients than would otherwise be detected. Strongly lensed SNe at $z \sim 1 - 2$ \citep{aman11,kelly14,glass,petr16} and lensed protogalaxies at $z \gtrsim 6$ \citep[e.g.][]{zhe12,coe13,brd14,vanz14,ryd15,ryd17a} have already been found in studies of individual, well-resolved cluster lenses. Recent studies suggest that such surveys could detect SNe at even higher redshifts \citep[$z \gtrsim$ 10;][]{pan12a,wet13c}.  \citet{om10} calculated the number of lensed SNe that might be found by LSST at $z < 3.75$, and \citet{gn17} explain how this number could be increased by an order of magnitude \citep[see also][]{gng18}. 

In this paper we estimate the number of a variety of strongly-lensed SNe that could be found by LSST at $z = $ 5 - 7. Although LSST optical bands limit SN detections to this redshift, they could constrain cosmic SFRs at the epoch of reionization, and a few of them might even be Pop III SNe, given that numerical simulations predict the formation of zero-metallicity stars down to $z \sim$ 3 \citep{tss09} and large parcels of pristine gas have now been discovered at $z \sim$ 2 \citep{fop11}. Lensed supernovae of any type at these redshifts may also be interesting as probes of the cosmic expansion history \citep{fm18}. In Section~\ref{sec:method} we discuss our SN spectra and explosion rates. We also describe how magnification maps are calculated for large fields and combined with redshifted spectra in specific filters to obtain SN detection rates.  Event rates for the LSST main survey and a proposed deep survey optimized to detect Pop III explosions  are examined in Section~\ref{sec:results} and we conclude in Section~\ref{sec:conclusion}. 

\section{Method}
\label{sec:method}

We convolve cosmic SN rates (SNRs) as a function of redshift with spectra from radiation hydrodynamical simulations, new magnification maps for wide fields and LSST filter response functions to predict detection rates for a variety of lensed SNe.  We assume $\Lambda$CDM cosmological parameters from the first-year \textit{Planck} best fit lowP+lensing+BAO+JLA+H$_0$: H$_0 = 67.3$, $\Omega_\mathrm{M}=0.308$, and $\Omega_{\Lambda}=0.692$  \citep{planck}.  We have verified that these parameters yield SN detection rates that are essentially identical to those obtained with the more recent \cite{planck2} parameters.

\subsection{SN Spectra}
\label{sec:sn_types}

We use Pop~III SNe as proxies for explosions at all metallicities for simplicity.  At $z=$ 5 - 7 most stars will be enriched to some degree, but the computational costs of a grid of spectra in both progenitor mass and metallicity would be prohibitive.  Furthermore, the average global metallicities one would assign to stars at any given redshift in this range are not yet known. Although 10~-~30~\Ms\ Pop~III stars are expected to be hotter and more compact than Pop~II/I stars of equal mass because of their lower internal opacities \citep{tgs01}, they are expected to explode with similar energies and luminosities (see \citealt{cl04,wh07} and Figure 1 of \citealt{wf12}).  However, the use of Pop~III CC SNe as substitutes for Pop~II/I explosions could cause us to underestimate their rest frame luminosities, as  $U$, $B$, $V$, $R$ and $I$ band light curves for our 15 \Ms\ 1.2 foe RSG explosion are about ten times dimmer than those for the M15\_E1.2\_Z1 model in \citet{kw09} (1 foe $= 10^{51}$ erg; see Figure 7).  We therefore also calculate, as an upper limiting case, LSST detection rates assuming our CC SN spectra to be a factor 10 brighter.

\subsubsection{Final Fates of Pop III Stars}
 
Non-rotating 8~-~30~\Ms\ Pop III stars die as core-collapse (CC) SNe and 140~-~260~\Ms\ stars explode as pair-instability (PI) SNe (\citealt{hw02}; see also \citealt{rs67,brk67,jw11,chen14c}).  30~-~90~\Ms\ stars collapse to BHs unless they are very rapidly rotating, in which case they can produce a gamma-ray burst \citep[GRB; e.g.,][]{gou04,yoon05,bl06b,wet08c,met12a,mes13a} or a hypernova \citep[HN;][]{nom10}.  Stars more massive than 260~\Ms\ encounter the photodisintegration instability and collapse to BHs, but at very high masses ($\gtrsim$~50~000~\Ms) a few stars may die as extremely energetic SNe due to the general relativistic instability \citep{montero12,jet13a,wet13a,wet13b,wet12d,chen14b}.

Pop III stars can actually encounter the PI at masses as low as $\sim$ 100 \Ms, but below 140 \Ms\ it triggers the ejection of multiple, massive shells rather than the complete destruction of the star. Collisions between these shells can then produce very bright events in the UV \citep[pulsational PI (PPI) SNe;][]{wbh07,cooke12,cw12a,chen14a}.  Rotation can cause stars to explode as PI SNe at masses as low as $\sim$~85~\Ms\ (rotational PI (RPI) SNe; \citealt{cw12,cwc13} -- see also \citealt{yoon12}).  Very massive Pop II/I stars with metallicities below $\sim$ 0.3 \Zs\ can retain enough mass to die as PI SNe \citep{lang07,wet13e,kz14b}.  10 ~-~ 30~\Ms\ stars can also eject shells prior to explosion, and the collision of the ejecta with the shell can, like PPI SNe, produce highly luminous events that are brighter than the explosion itself (Type IIn SNe).

In our study we include spectra for 150, 175, 200, 225 and 250~\Ms\ Pop III PI SNe (\citealt{wet12a,wet12b}; see \citealt{kasen11,det12,kz14a,jer16} for other PI SN spectra and light curves). We also have spectra for 15 and 25 \Ms\ Pop III CC SNe \citep{wet12c}, each of which can have explosion energies of 0.6, 1.2 or 2.4 foe with equal probablility. The structure of the star at death can have a profound effect on the light curve of the explosion but is not well constrained by 1D evolution models, so we consider explosions of both compact blue supergiants (BSGs) and red supergiants (RSGs) at each mass to bracket the range of spectra expected for CC and PI SNe. Consistent with the numbers of Type IIP SNe observed today, we take 75\% of the stars to die as RSGs.  

We also include spectra for 40 \Ms\ Pop III Type IIn SNe \citep{wet12e}, assuming that 1\% of all CC SNe produce shell collisions, but ignore GRBs and HNe because of their small numbers ($\lesssim$ 10$^{-3}$ of all CC events). Our grid of models also has spectra for 90 - 140 \Ms\ RPI SNe at 5 \Ms\ increments, which are essentially the explosions of stripped He cores (\citealt{smidt14a}; see \citealt{cw14a} for additional spectra). A 110~\Ms\ PPI SN is also included \citep{wet13d}. More or less massive PPI progenitors either do not produce collisions that are bright in the visible/NIR today or shells that collide at all. Since this mass falls in the same range as RPI SNe we assume that 50\% of the stars from $107.5-112.5$ \Ms\ die as PPI SNe.

Primordial stars are not thought to lose much mass over their lives because of their low internal opacities, which prevent strong winds \citep{bhw01,kud00,vink01,Ekstr08}.  Consequently, their explosion rates can be derived directly from their SFRs and IMF because their masses at death are nearly those at birth.  But mass loss in Pop I and II stars can change some of them into other mass ranges in which they will explode at the end of their lives while removing others.  This introduces some ambiguity into their explosion rates because mass loss rates are not well constrained by either stellar mass or metallicity. For simplicity, we make no attempt to account for mass loss in our Pop II/I SNRs and just apply the same mass ranges for Pop III SNe to them.

\subsubsection{Spectrum Models}
\label{sec:popIIISNspectra}

Our spectra were calculated in three stages. First, the progenitor star was evolved from birth to the onset of collapse and explosion in Kepler \citep{kep1,kep2} or GENEVA \citep{e08,hle13,hle16}, which are one-dimensional (1D) Lagrangian stellar evolution codes.  After all explosive burning was complete (typically in 10 - 30 sec) the shock, surrounding star, and its envelope were mapped onto a 1D Eulerian grid in the radiation hydrodynamics code RAGE \citep{rage} and evolved for 6 months to 3 years, depending on explosion type. Blast profiles from RAGE were then post processed with the SPECTRUM code \citep{fet12} to obtain source frame spectra for the fireball at every stage of its evolution.  

Flux-limited diffusion (FLD) with gray OPLIB opacities \citep{oplib} in RAGE were used to transport radiation through the SN ejecta, but monochromatic OPLIB opacities were used in SPECTRUM to produce spectra with 13899 wavelengths from hard X-rays down to the far IR.  Gray FLD enables RAGE to properly capture shock breakout from the star and the expansion of the ejecta into the IGM while monochromatic opacities allow SPECTRUM to synthesize emission and absorption lines from the fireball in addition to its continuum.  The RAGE simulations were performed with 50~000~-~200~000 zones and $2 - 5$ levels of AMR refinement. More details on the physics and setup of the models can be found in their respective papers, cited above.

\subsection{AB Magnitudes}
\label{sec:lightcurves}

SN light curves are usually characterized by a brief, intense pulse associated with shock breakout from the surface of the star followed by either a steep or gradual decline as the fireball expands and cools.  The light curve may rebrighten at later times if the explosion creates large masses of \Ni, whose decay photons diffuse out of the ejecta on timescales of weeks to months. To compute AB magnitudes for a SN in a specific filter we first cosmologically redshift and dim its rest-frame spectrum, $F'(\lambda)$:
\begin{equation}
F(\lambda) = \frac{F'\left(\frac{\lambda}{1+z}\right)}{(1+z) 4 \pi d^2_{\mathrm{L}}(z)}.
\end{equation}
Here, $d_{\mathrm{L}}(z)$ is the luminosity distance: 
\begin{equation}
d_{\mathrm{L}}(z) = (1+z) c/\mathrm{H}_0 \int_0^z \frac{1}{\sqrt{\Omega_{\mathrm{M}} (1+z)^3 + \Omega_{\lambda}}} dz.
\end{equation}
AB magnitudes, $m_{\mathrm{AB}}$, in specific filters are then computed from
\begin{equation}
m_{\mathrm{AB}} = -2.5 \, \mathrm{log_{10}} \left( \frac{\int_0^{\infty} F(\lambda) T(\lambda) d \lambda}{\int_0^{\infty} F_0(\lambda) T(\lambda) d \lambda} \right), \label{eq:magnitudecalculation}
\end{equation}
where $T(\lambda)$ is the filter transmission function and $F_0(\lambda) = 3.630781 \times 10^{-20} c \lambda^{-2}~\mathrm{ergs}~\mathrm{cm}^{-2}~\mathrm{s}^{-1}~\mathrm{m}^{-1}$, the reference spectrum for AB magnitudes, $c$ standing for the speed of light.

Radiation blueward of $\lambda_{\mathrm{Ly\alpha}}=1216$~\AA{} emitted at $z >$ 6 may be redshifted into $\lambda_{\mathrm{Ly\alpha}}$, be resonantly scattered, and never be observed \citep{gp65}. Consequently, at $z>6$ we simply take all SN flux shorter than $\lambda_{\mathrm{Ly\alpha}}$ to be absorbed by the IGM and remove it from the spectrum. After reionization, there still exist clouds of neutral hydrogen so some flux blueward of $\lambda_{\mathrm{Ly\alpha}}$ from events at $z <$ 6 may still be absorbed. To account for this partial absorption in our spectra we use the model in \citet{madau95}.

\subsection{Cosmic SFRs/SNRs}
\label{sec:sfr}

Cosmic SNRs depend directly on global SFRs and their mass functions. SFRs derived from semianalytical models \citep[e.g.,][]{wa05,wl05} and numerical simulations over the years have varied by a factor of 200 or more depending on redshift \citep[see, e.g., Figure 5 of][]{wet13c}. More recently, simulations using older cosmological parameters have also produced SFRs that are not consistent with current constraints on the optical depth to Thomson scattering by free electrons at high $z$ \citep[$\tau_{\mathrm{e}}$;][]{vis15}. To avoid these difficulties we compute Pop~II SNRs from cosmic SFRs extrapolated from observations of GRBs, which are used as tracers of early SF \citep{re12}. We apply the Salpeter IMF \citep{sal55} to convert these rates to SNRs, adopting the lower limits to the GRB rates to be conservative.

Pop III SNRs were calculated with a semi-analytical model that utilises a cosmologically representative set of halo merger trees with detailed prescriptions for radiative and chemical feedback on star formation in the halos \citep{til15a,magg16}.  This model assumes a logarithmically flat IMF \citep{get11} for Pop III stars, whose masses are randomly sampled over an interval of 1 - 300 \Ms. This IMF was chosen because it produces the full range of explosions expected for primordial stars. By tracing all the individual randomly generated Pop~III stars, the model makes self-consistent predictions for SNRs over the mass range of each explosion type discussed earlier. 

\begin{figure}
\begin{center}
\includegraphics[width=80mm]{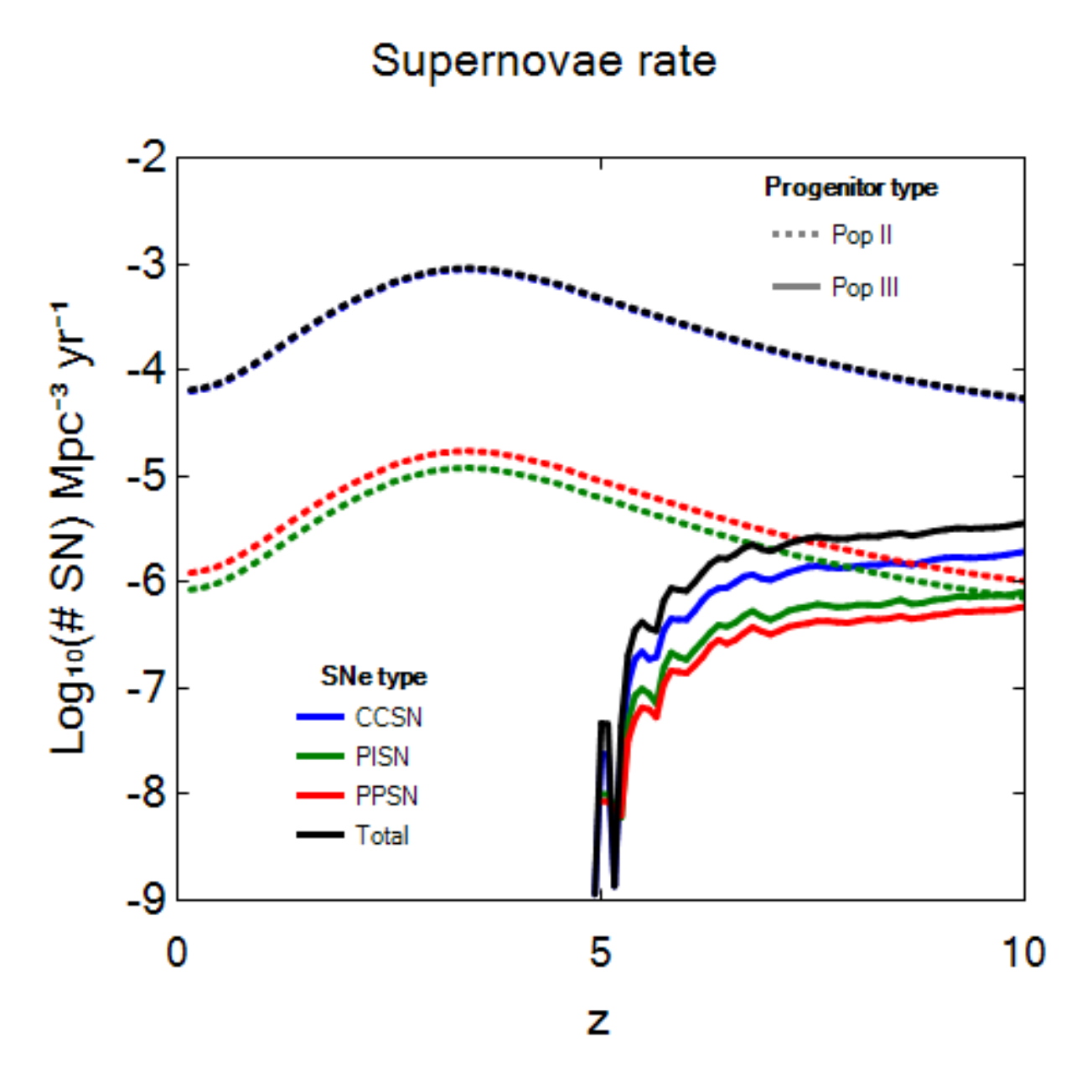}
\end{center}
\caption{Pop III (solid) and Pop II (dashed) SNRs as a function of redshift, including CC (blue), PI (green), RPI plus PPI (red) and total (black) SNRs. The Pop~II CC SNRs are essentially indistiguishable from the total Pop~II rates.}
\label{fig:semianalytical}
\end{figure}

SNRs within each explosion type are further partitioned by progenitor masses for which there are light curves.  Thus, CC SNRs are divided into two mass bins from 15 - 30 \Ms, RPI SNRs are partitioned across nine bins from 90 - 140 \Ms, and PI SNRs are calculated for five bins from 150 - 250 \Ms.  We show total SNRs for each explosion type as a function of redshift in Figure~\ref{fig:semianalytical}.  As expected, Pop III SN rates taper off with redshift as the universe becomes more chemically enriched, falling off dramatically at z~$\sim$~5 when metals become present in most of the universe.  In contrast, Pop~II SNRs rise over this interval as metals contaminate more and more stars, reaching a peak at $z \sim$~2~-~3 that coincides with the observed peak in cosmic SFR due to galaxy formation.

\subsection{Magnification Functions}
\label{sec:mu}

We calculate magnification functions in wide fields due to strong lensing by both galaxy clusters and massive galaxies. These functions, or maps, predict what fraction of the sky is magnified by a factor of $\mu$ or more as a function of source redshift. Magnifcation maps for galaxy clusters are derived from a statistical model developed by \citet{mm14}. This model takes the lensing objects to be dark matter haloes with Navarro-Frenk-White (NFW) density profiles \citet{nfw96} with a Sheth-Tormen distribution in mass, $N(m,z)$, where $N$ is the number density of haloes at or above a mass $m$ at redshift $z$. For each redshift, the area on the sky that is magnified by a factor of $\mu$ or more is calculated by adding the contribution from every object at lower redshifts. $P_{\mathrm{S}}(\mu, z_{\mathrm{S}})$, the fraction of the area in the source plane at redshift $z_{\mathrm{S}}$ with magnification greater than $\mu$, is the optical depth of all lenses between the observer and the source: 
\begin{equation}
        P_{\mathrm{S}}(\mu, z_{\mathrm{S}}) = \frac{1}{4\pi D_s^2}\int_0^{z_s}\int_0^\infty  
N(m,z)\sigma_d(m,z,z_s) \mbox{d}m \mbox{d}z,
\end{equation}
where $D_s$ is the angular diameter distance to the source and
\begin{equation}
   \sigma_d(z_s) = \eta_0^2\int_{B_l} \frac{\mbox{d}^2x}{|\mu(x)|}
\end{equation}
is the cross section of a lens with a length-to-width ratio greater than $d$ over an area $B_l$ in the lens plane. Here we define the dimensionless coordinates $x=\theta/\theta_0$ and $\eta_0=(D_s/D_l)\theta$ for the lens and source planes, respectively. 

We take the lenses to be distributed sparsely enough that their areas do not overlap and are thus additive. These magnification areas become less certain at low $\mu$ because NFW profiles more accurately represent lensing objects near their cores where densities are high. Our assumption that lensing areas do not overlap also breaks down at low $\mu$, so we only compute magnification areas for $\mu \geqslant 2$. For 1 $< \mu <$ 2 we logarithmically extrapolate the area from its value at $\mu =$ 2 (Section~\ref{sec:disc}). This approach does not return the original field of view (FoV) as $\mu$ approaches 1 so our maps represent a lower limit to the area on the sky that is strongly lensed. At $\mu = 1$ we simply take the magnified area to be the FoV.  At the other extreme, the magnified area falls as $\mu^{-2}$ at large $\mu$ for geometrical reasons, so at $\mu >$ 100 we take it to be the area at $\mu =$ 100 scaled downward by this factor.

We calculate magnification functions for massive galaxies with the \citep{col15} model. The galaxies are assumed to be isothermal ellipsoids with velocity dispersions drawn from the observed velocity dispersion function from SDSS \citep{cpv07}. We truncate the velocity dispersion function between 50 and 400~km~s$^{-1}$. Like clusters, these lensing galaxies are assumed to be randomly distributed but they are taken to have a constant comoving density out to $z=2$. We draw lens ellipticities from the velocity dispersion dependent fit of Equation 4 in \citet{col15} and have neglected the effect of compound lensing, which, while rare, can produce extreme magnifications of high redshift sources \citep{cb16}. The parameters for both models are listed in Table~\ref{tab:mags}. The mass limits listed for the  \citet{col15} model are the masses residing within the Einstein radius. Since the two models cover different ranges in lens mass, the total area that is magnified within the FoV can be approximated by simply summing their respective contributions. The \citet{col15} model produces magnified areas that are somewhat larger than those of \citet{mm14}, by up to 50\%, but their overall agreement suggests that our areas are accurate to within an order of magnitude.  Rather than sum the respective $P_{\mathrm{S}}(\mu, z_{\mathrm{S}})$ of the two models to obtain the total area of the FoV that is lensed, we calculate the number of lensed SNe from each model separately and sum them to obtain the total number which is equivalent.

\begin{table}
\caption{Magnification Model Parameters}
\label{tab:mags}
\begin{center}
\begin{tabular}{lll}
\hline
\\
& Maturi & Collett \\
\hline
\\
Mass distribution & Sheth-Tormen & Empirical \\
Density profile & NFW  & Isothermal ellipsoid \\
Mass range & $10^{13}$ - $10^{15}$~\Ms/h & $< 10^{13}$ \Ms\ \\
Redshift range & $0.05 < z < 10$ & $0.015 < z < 2.0$ \\
\\
\hline
\end{tabular}
\end{center}
\end{table}

\subsection{Lensed SN Detection Rates}

The magnification required to detect a SN at redshift $z$ is
\begin{equation}
\mu(\tau, z, \mathrm{F}(\tau)) = \mathrm{max}(10^{0.4 (m(\tau, z) - m_{\mathrm{L}})},1),
\end{equation}
where $\mathrm{F}(\tau)$ is the spectrum at a time $\tau$ after shock breakout in the observer frame, $m(\tau, z)$ is the AB magnitude for the SN, and $m_{\mathrm{L}}$ is the AB magnitude detection limit of the survey in the given filter. This minimum magnification is then used to compute $P_{\mathrm{S}}(\mu, z_{\mathrm{S}})$ to calculate the area in the source plane with enough magnification to detect the SN. We neglect demagnification by cosmic voids \citep{hilb09,mason15}.

The total number of lensed SNe observed in an interval $\Delta z$ at $z$ is
\begin{equation}
N(z,\Delta z) = \int^{z + \Delta z}_{z} S(z) A_{\mathrm{FoV}} dV(z),\label{eq:numbercount}
\end{equation}
where $A_{\mathrm{FoV}}$ is the angular area covered by the survey or FoV of the telescope and 
\begin{equation}
S(z) = \int^{\tau_{\mathrm{U}}}_{0} P_{\mathrm{S}}(\mu(\tau, z, \mathrm{F}(\tau)), z_{\mathrm{S}}) \mathrm{SNR}(z) d\tau.\label{eq:sz}
\end{equation}
the total number of SNe in a mass range per volume (in Mpc$^{-3}$) that are visible at $z$. Here, SNR$(z)$ is the supernova rate in Mpc$^{-3}$ yr$^{-1}$, $\tau_{\mathrm{U}}$ is the upper limit in ages for which we have spectra, and the integration is performed over $\tau$ to capture SNe of all ages. We assume that $\tau_{\mathrm{U}}$ is small in comparison to the cosmological time corresponding to $\Delta z$ since an event visible at $z$ actually exploded at a slightly higher $z$. To predict the number of lensed SNe of a particular type that would be detected,  $N(z,\Delta z)$ is summed over the corresponding range in progenitor mass.

\subsection{Transient Identification}
\label{sec:identifying_detections}

$N(z,\Delta z)$ is the number of SNe that would appear in a single or stacked observation of the FoV under the assumption that their luminosities remain nearly constant over the exposure time. But they can only be identified as SNe by subtracting two or more exposures of the same FoV, a procedure we call image pairing. Objects that appear, disappear, or change in brightness over the time between observations, or cadence $\Delta \tau$, are tagged as SN candidates. To verify that a candidate actually is a SN, follow-up observations are required. Ideally, spectroscopic measurements of distinctive emission lines could identify the transient as a SN. Alternatively, broadband follow-up observations could match the light curve to a certain SN type. To perform such followups the event must be visible after initial identification, so we discard transients disappearing between observations. Our SN detections are therefore objects that appear between exposures as well as objects that have changed brightness by more than  25\%. As this is an arbitrary threshold, chosen to reflect changes in flux that can be easily detected, we considered a range of thresholds to investigate their effect on our number counts in Section~\ref{sec:disc}.

Variable stars in our galaxy can appear as SN contaminants in image pairings unless they are known and catalogued. However, even previously unknown variable stars will only contaminate the first pairing and, after being identified as such, can be discarded from subsequent pairings. Spectroscopic follow-up could distinguish SNe from variable stars. Alternatively, new detections at the same position in later observations would likewise reveal the object to be a variable star.

\subsubsection{New SNe}

If P$_{\mathrm{S}}(\mu, z_{\mathrm{S}})$ is the fractional area of a FoV in which a lensed SN is visible at a given stage of evolution, then P$_{\mathrm{S}}(\mu, z_{\mathrm{S}})=1$ if m($\tau, z) < m_{\mathrm{L}}$. If the fractions of the FoV in which the SN is visible in the first and second images of a pairing are P$_{\mathrm{S}}(\mu_1, z_{\mathrm{S}})$ and P$_{\mathrm{S}}(\mu_2, z_{\mathrm{S}})$, where $\mu_1$ and $\mu_2$ are the magnifications required to observe the SN in images 1 and 2 respectively, then the fractional area in which the SN appears in the pairing itself is 
\begin{equation}
\Delta P_{\mathrm{S}}(z_{\mathrm{S}}) = P_{\mathrm{S}}(\mu_2, z_{\mathrm{S}}) - P_{\mathrm{S}}(\mu_1, z_{\mathrm{S}})~\mathrm{if}~P_{\mathrm{S}}(\mu_2, z_{\mathrm{S}}) > P_{\mathrm{S}}(\mu_1, z_{\mathrm{S}}), \label{eq:Psz}
\end{equation}
where $\mu_1=\mu(\tau-\Delta\tau/(1+z))$ and $\mu_2=\mu(\tau)$. $\Delta P_{\mathrm{S}}(z_{\mathrm{S}})$ is then used in Equation~\ref{eq:sz} as $P_{\mathrm{S}}(\mu(\tau, z, \mathrm{F}(\tau)), z_{\mathrm{S}})$.

\subsubsection{Evolving SN}
\label{sec:changingSN}

We flag a change in flux in the FoV as a SN if it appears in consecutive images and it has increased or decreased by 25\% or more. Since the candidate appears in both images the image with the smallest area with enough magnification for a detection should be used: 
\begin{equation}
P_{\mathrm{S}}(z_{\mathrm{S}}) = \mathrm{min}(P_{\mathrm{S}}(\mu_1, z_{\mathrm{S}}), P_{\mathrm{S}}(\mu_2, z_{\mathrm{S}})).\label{eq:PsChangez}
\end{equation}
We also require that $|m(\tau-\Delta\tau/(1+z), z) - m(\tau, z)| \geq 2.5 \, \mathrm{log}_{10}(1.25)$, which guarantees that the brightness has changed by at least 25\% or more. As in Equation~\ref{eq:Psz}, $\mu_1=\mu(\tau-\Delta\tau/(1+z))$ and $\mu_2=\mu(\tau)$. $P_{\mathrm{S}}(z_{\mathrm{S}})$ is then used in Equation~\ref{eq:sz}.

\subsubsection{Multiple Flaggings of the Same SN}
\label{sec:consecutiveobservations}

In principle, a SN can explode, fade below $m_{\mathrm{L}}$, and then later rebrighten and  become observable again. Such events could be mistakenly flagged as new SNe twice in different image-pairings.  When considering the area in which new SNe are detected in consecutive exposures, all SNe that have already been observed should be discarded. $P_{\mathrm{S}}(\mu_1, z_{\mathrm{S}})$ must therefore be replaced by the largest $P_{\mathrm{S}}(\mu, z_{\mathrm{S}})$ from any previous image among the consecutive images. Equation~\ref{eq:Psz} becomes
\begin{equation}
\Delta P_{\mathrm{S}}(i, z_{\mathrm{S}}) = P_{\mathrm{S}}(\mu_i, z_{\mathrm{S}}) - \mathrm{max}_{k<i}(P_{\mathrm{S}}(\mu_k, z_{\mathrm{S}}))
\label{eq:Pszi}
\end{equation}
if $P_{\mathrm{S}}(\mu_i, z_{\mathrm{S}}) > \mathrm{max}_{k<i}(P_{\mathrm{S}}(\mu_k, z_{\mathrm{S}}))$.  Each $P_{\mathrm{S}}(\mu_i, z_{\mathrm{S}})$ corresponds to the fraction of the sky in which the SN can be detected in a given observation. The $i$ label the observations, i.e., $P_{\mathrm{S}}(\mu_{i+1}, z_{\mathrm{S}})$ is for an observation one cadence later. Since $P_{\mathrm{S}}(\mu_i, z_{\mathrm{S}})$ is bounded there is an $i_\mathrm{max}$ for which $P_{\mathrm{S}}(\mu_i, z_{\mathrm{S}})$ is maximum. The total area $P_{\mathrm{S}}(z_{\mathrm{S}})$ from all $i > i_\mathrm{max}$ is then zero. When summing over all $i < i_{\mathrm{max}}$ in Equation 11 we keep only the subset of $i$ for which $P_\mathrm{S}(\mu_i, z_s) > \mathrm{max}_{k<i}(P_\mathrm{S}(\mu_k, z_s))$. This subset retains all positive results from Equation 11. We index this subset with $m$, and Equation 11 then becomes
\begin{equation}
\Delta P_\mathrm{S}(m,z_\mathrm{S}) = P_\mathrm{S}(\mu_m,z_\mathrm{S})-P_\mathrm{S}(\mu_{m-1},z_\mathrm{S}). \label{eq:Ps}
\end{equation}
We sum this to obtain the fraction of the FoV in which it is possible to observe new, distinct SNe. Since before the explosion the area is 0 (i.e., $P_\mathrm{S}(\mu_0,s_\mathrm{S}) = 0$) there is a sum of disjunct, uninterrupted intervals from 0 to $\mathrm{max}(P_\mathrm{S}(\mu_i,z_\mathrm{S}))$ so the sum reduces to $\mathrm{max}(P_\mathrm{S}(\mu_i,z_\mathrm{S}))$. The number of SNe observed also depends on the SNR and $\Delta t$ between observations, where $\Delta t < \Delta \tau$ and is preferably as small as possible. To include the entire light curve in the calculation we add an additional index $n$ dividing $\Delta \tau$ in $N = \Delta \tau / \Delta t$ equal intervals. Equation~\ref{eq:Ps} is then applied to each interval, where $\mu_i = \mu_{i, n}$ denotes the iteration within $\Delta \tau$. The total fraction of the FoV magnified above the threshold for detection multiplied by the formation time is then
\begin{equation}
\sum_{n=1}^N P_{\mathrm{S, n}}(z_{\mathrm{S}}) \Delta t = \sum_{n=1}^N \mathrm{max} (P_{\mathrm{S}}(\mu_{i, n}, z_{\mathrm{S}})) \Delta t.
\end{equation}
Since in this fraction of the FoV usually just the peak of the light curve can be seen we call it the peak area. It does not take into account the start and end of the observations. At these points there might be observations of SNe part of the way through their light curves which might give rise to detections of unique SNe even though it would have been an extra detection of an already observed SNe if conducted later in the survey. This should, however, be a small effect and if the time for the consecutive observations is significantly longer than the duration of the light curves the effect will be negligible. In multiple consecutive observations, every evolving SN has been already counted as an emergent SN at some point, and the method only counts distinct events. 

\begin{table}
\caption{
The brightest CC SN and PI SN in the $z$ and $y$ bands. Here, $z$ is the redshift at which the SN is brightest in the given filter, m$_{\mathrm{AB}}$ is the peak AB magnitude of the event at this redshift, $t_{\mathrm{P}}$ is the time in the rest frame at which the SN reaches peak brightness, and $\mu_{\mathrm{min}}$ is the minimum magnification for the transient to be detected in the LSST main survey.
}
\label{tab:mostluminous}
\begin{center}
\begin{tabular}{lllll}
\hline
\\
Type & CC & PI & CC  & PI \\
\hline
\\
Filter & $z$  & $z$ & $y$ & $y$ \\
$M_{\star}$ (\Ms) & 15  & 250  & 15  & 250  \\
$E_{\mathrm{SN}}$ (foe) & 2.4 & 94 & 2.4 & 94 \\
$z$ & 5.5 & 5.0 & 5.1  & 5.0 \\
$t_{\mathrm{P}}$ & 0.8~d  & 21.2~d & 5.4~d & 21.2~d \\
m$_{\mathrm{AB}}$ & 29.36 & 25.90 & 29.21 & 25.95 \\
$\mu_{\mathrm{min}}$ & 260 & 11 & 510 & 25 \\
\\
\hline
\end{tabular}
\end{center}
\end{table}

\begin{figure*}
\begin{center}
\includegraphics[width=180mm]{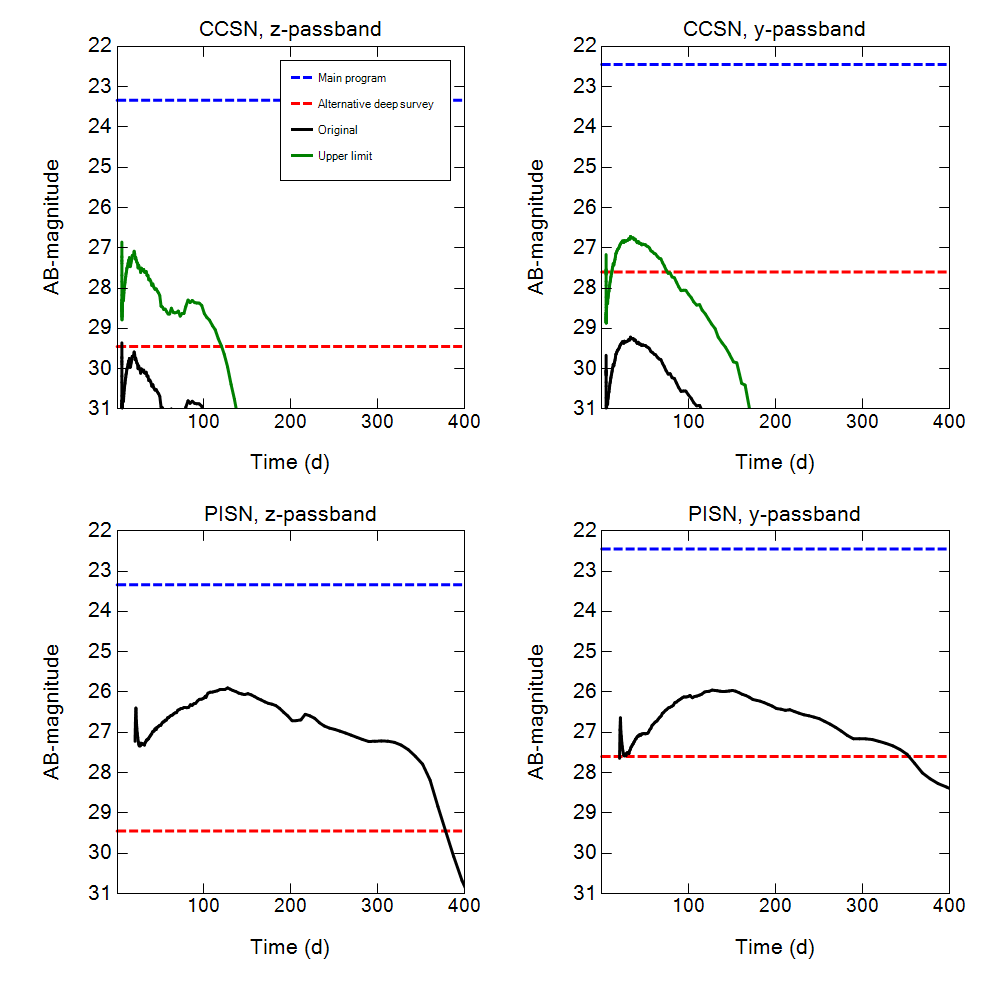}
\end{center}
\caption{
Light curves for the brightest CC SN (top panels) and PI SN (bottom panels) in the $z$ and $y$ bands at $z =$ 5.1 and 5.5, respectively. The x-axis is the time in the observer frame $\tau$. In the upper panels the black line is the Pop~III CC SN light curve and the green line is the light curve expected for a Pop~I CC SN of equal energy and progenitor mass, which is  brighter as discussed in Section~\ref{sec:popIIISNspectra}. The dashed horizontal lines are the detection limits for the main survey (blue) and the ADS (red).
}
\label{fig:lightcurve}
\end{figure*}

\section{Results}
\label{sec:results}

We consider three surveys. The first is the main survey, or Wide Deep Fast survey (WDF). The second is a 10 deg$^2$ 'deep-drilling' (DD) survey for which we determine the AB magnitude limits required to detect a given type of SN at high redshift. The third is an alternative deep survey (ADS) whose purpose is to detect Pop III SNe. It consists of two deeper exposures one year apart, and we vary the area while holding the exposure time constant to determine the field that would maximize the number of Pop~III SN counts, as described later.

The LSST $u$, $g$, $r$, $i$, $z$ and $y$ bands cover $3300-3900$~\AA{}, $4100-5400$~\AA{}, $5600-6900$~\AA{}, $6800-8400$~\AA{}, $8000-9350$~\AA{} and $9000-10300$~\AA{}, respectively. As discussed in Section~\ref{sec:introduction}, the Gunn-Peterson trough imposes upper limits on the redshift at which an event can be observed in a given filter. For the $u$, $g$, $r$, $i$, $z$, and $y$ filters these limits are $z =$ 2.3, 3.5, 4.7, 6.0, 6.7, and 7.5, respectively. We exclude the $u$, $g$ and $r$ passbands from our study because their redshift limits lie below $z = 5$ and only consider the $y$ and $z$ bands. 

Table~\ref{tab:mostluminous} shows some of the properties of the CC SN and PI SN that are brightest in the $y$ and $z$ bands. The CC SN is brightest in these filters at $z = 5.5$ and 5.1 rather than at the lowest redshift $z =$ 5, at which it is closest to Earth. This is due to the fact that more flux may be redshifted into a given filter from a higher $z$ if it originates from a brighter region of the rest frame spectrum, and this can more than compensate for the greater distance to the event.  The PI SN is brightest in both bands at the lowest redshift considered, $z =$ 5. We show light curves for these explosions in Figure~\ref{fig:lightcurve}, including the upper limiting case of a Pop~I CC SN that is 2 - 3 magnitudes brighter than a Pop III CC SN of equal energy. None of the SNe reach the detection limit of the WDF so they would have to be strongly lensed to be observed. The PI SN is visible without lensing in the ADS but only the brighter upper limiting case of CC SN is visible in the ADS without it.

\subsection{WDF}
\label{sec:mainprogram}

The WDF will reach m$_{\mathrm{AB}}$= 23.3 and 22.5 in the $z$ and $y$ bands, respectively, for 30s exposures \citep[LSST takes pairs of 15s exposures each night;][]{LSST}. The exposures range over 30000~deg$^2$ on the sky for an actual area of 18000~deg$^2$, which we take to be the FoV. Each location on the sky will be visited about 800 times over 10~years for an average revisit every 27 days in each filter. Although cadences from a few days up to 10 years can be obtained with the proper choice of images, we adopt 27 days as the cadence in our study. Number counts for new SNe are maximized by using as short a cadence as possible since some SNe might appear and then fade below the detection limit of the survey at longer cadences.  Also, fewer images can be used at longer cadences unless they overlap.

\begin{figure}
\begin{center}
\includegraphics[width=80mm]{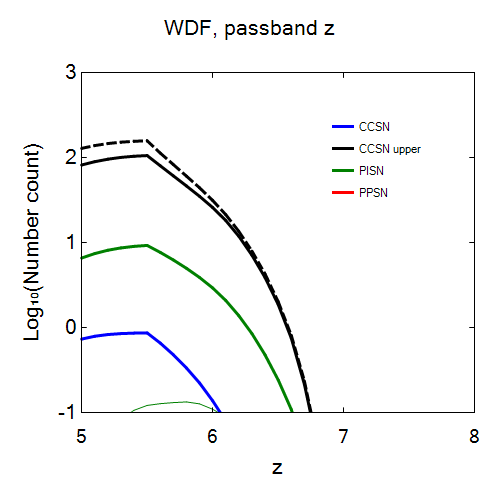}
\end{center}
\caption{
Number of Pop~II and Pop~III SNe detected per unit redshift in the WDF survey (thick and thin lines, respectively). These SNe are all lensed because the WDF lacks the sensitivity to detect unlensed events.  The black dashed line is the predicted number of all explosions when SNe that rebrighten are not rejected as new events. The failure to properly identify such SNe can lead to number counts that are nearly twice the actual values. 
}
\label{fig:mainprogramresults_z}
\end{figure}

\begin{table}
\caption{
Predicted total counts of lensed SNe for the WDF survey from 5~$<$~$z$~$<$~7 by filter and type.
}
\label{tab:MS}
\begin{center}
\begin{tabular}{llll}
\hline
\\
			&		& Filter	& 	 \\
SN Type 		& $i$  	& $z$ 	& $y$ \\
\hline
\\
Pop II CC (hi)	& \popIIccihigh{}	& \popIIcczhigh{}	& \popIIccyhigh{}  \\
Pop II CC (lo)	& \popIIccilow{}		& \popIIcczlow{}		& \popIIccylow{}    \\
Pop II PI		& \popIIpii{}		& \popIIpiz{}	& \popIIpiy{} \\
Pop II IIn		& 0		& \popIInz{}		& 0	 \\
Pop II PPI 	& 0 		& \popIIppiz{}	& 0    \\
Pop III CC (hi) 	& 0  		& 0		& 0    \\
Pop III CC (lo) 	& 0 		& 0		& 0    \\
Pop III PI 		& 0 		& \popIIIpiz{}		& 0    \\
Pop III IIn 		& 0 		& 0 		& 0    \\
Pop III PPI 	& 0 		& 0 		& 0    \\
\\
\hline
\end{tabular}
\end{center}
\end{table}

\begin{figure}
\begin{center}
\includegraphics[width=80mm]{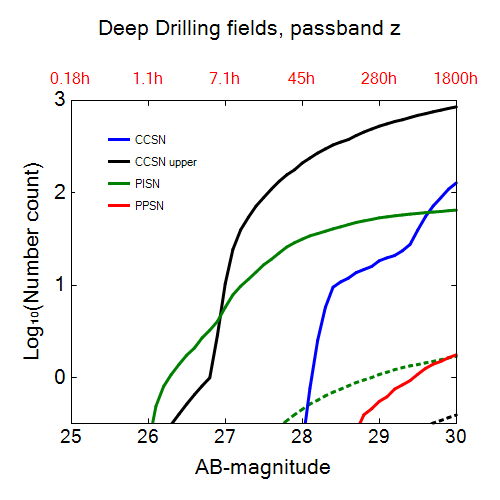}
\end{center}
\caption{
Total number of Pop II and Pop III SNe detected from $5 < z < 7$ as a function of observational depth in AB magnitude in the DD survey (solid and dotted lines, respectively). Most of these events are not lensed because of the low probability of lensing in the relatively small FoV. The few that are lensed can be seen in the CC SN upper plot as the break in slope at $z \sim$~27. The break in slope in the blue CC SN detections at $z \sim$ 29 is due to Type IIn SNe. On the upper axis is the exposure time necessary to reach the corresponding AB magnitude on the lower axis.
}
\label{fig:abmaggraph_z}
\end{figure}

As noted above, the WDF cannot detect unlensed SNe at $z \sim$~5~-~7 at all in the $u$, $g$ and $r$ bands and it is too shallow to find them in the $i$, $z$, or $y$ bands. However, as shown in Figure~\ref{fig:mainprogramresults_z} and Table~\ref{tab:MS}, it could find large numbers of lensed Pop~I/II SNe in the $z$-band: $\sim \popIIcczlow{}-\popIIcczhigh{}$ CC SNe depending on their brightness and $\sim$ \popIIpiz{} PI SNe, but no PPI SNe. Only a fraction of these events will appear in the $y$-band: $\sim$ \popIIpiy{} PI SNe, no PPI SNe, and at most  \popIIccyhigh{} CC SNe. No Pop~III SNe will appear in any of the bands due to the low numbers of Pop~III stars at $z < 7$, not the intrinsic brightness of their explosions.

It is evident from the black dashed line in Figure~\ref{fig:mainprogramresults_z} that the failure to reject rebrightening as a spurious event can lead to number counts that are nearly twice the actual values.  As noted in Section~\ref{sec:consecutiveobservations}, variations in the light curves of some SNe can cause them to be detected, fade, and then be detected again when they brighten later, mimicking two new events if they are not properly identified. The breaks in the plots at $z \sim$ 5.5 are due to Ly$\alpha$ beginning to be redshifted into the filter: at higher redshifts more and more of the radiation from the explosions is scattered by neutral hydrogen so detection rates fall.

\subsection{DD}

We show the total number of SN detections in a single 10~deg$^2$ FoV as a function of survey depth in AB magnitude in Figure~\ref{fig:abmaggraph_z}, assuming two observations one year apart. The exposure time required for a given magnitude on the lower axis is noted in red on the upper axis. There are larger uncertainties in SN counts at longer exposure times  because the luminosity of an event can vary over these times. This is less of an issue with PI SNe because their luminosities evolve more slowly than those of CC SNe, which are dimmer and more transient. Nevertheless, at the redshifts we consider they are still extended events in comparison to the 1 - 2 night explosion peaks of most low-redshift SNe. The timescales on which the light curves in Figure~\ref{fig:lightcurve} evolve suggest that exposure times can be no longer than $\sim$ 80h (10~days) for CC SNe and 800h (100~days) for PI SNe to obtain accurate counts.

The SN number counts are quite sensitive to survey depth. To detect Pop II CC SNe the DD field must reach AB magnitudes of 27 - 28, or exposure times of 7.1 - 45h. This range is derived from the upper and lower limits to the luminosities of CC SNe. To find Pop II PI SNe the DD field must slightly exceed AB mag 26,  or a few hours of exposure, while Pop II PPI SNe and Pop III PI SNe require AB magnitudes above 29, or more than 300h of exposure.

These numbers can be explained by Figure~\ref{fig:lightcurve}. Pop~II SNe become visible as soon as their brightest peek is observable. This is due to their sheer number causing a signicant count as soon as they are observable even a short time. Indeed, lensing makes it almost possible to observe lensed CC SN at even fainter magnitudes for the upper limits used as can be seen by the change of slope at $\sim$~27 AB-magnitudes. For the lower limits CCSNe the observational limit is pushed towards fainter magnitudes by the inclusion of Type~IIn which has a brighter peak. The lower rates for Pop~III SNe must be compensated by the possibility of observing the light curve for a more extended time period. Only by reaching as faint as AB-magnitude 29 will this be possible.

\subsection{ADS}
\label{sec:alternativedeepsurvey}

Two 640h (80 day) exposures in a single 10~deg$^2$ field one year apart in the ADS would yield a gain of 6.1 in AB magnitude over the WDF, m$_{\mathrm{AB}}=29.5$ in the $z$ band. Each exposure is assumed to be a single image in our calculations, and is long enough that the light curve of an event could significantly evolve over its duration, which could distort number counts. This is less of an issue for PI SNe because their luminosities evolve relatively slowly after peak but it is more important for CC SNe because they vary more rapidly. However, we use the single exposure approximation as a starting point. Then, in a separate calculation,  we vary the survey area while holding the total exposure time constant to determine what survey area and depth would maximize Pop III SN number counts. Larger survey areas here mean greater opportunities for lensing but shorter exposures per pointing, so this calculation explores the tradeoff between the two.

\begin{figure}
\begin{center}
\includegraphics[width=80mm]{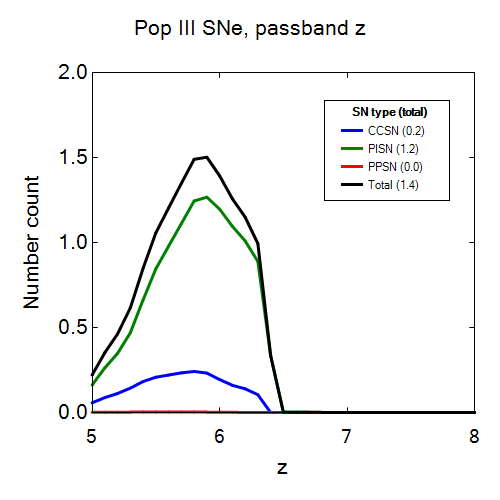}
\end{center}
\caption{
Pop~III SN number counts in the $z$ band for two 640h exposures in a single 10~deg$^2$ FoV one year apart in the ADS. These are all unlensed events because of the low probability of lensing in the relatively small FoV. The most likely detections are of PI SNe at $z < 6.4$. It is unlikely that any Pop~III CC SNe will be detected. On average about one SN is  expected to be found at $z \sim 6$.
}
\label{fig:ADS}
\end{figure}

We show number counts for Pop III SNe by explosion type for the simplest case in which the 640h exposures are carried out in a single 10 deg$^2$ field in Figure~\ref{fig:ADS}. About 1 unlensed PI SN is expected from $z \sim$ 5 - 6.4 in the $z$ band. CC SN detections are unlikely even at their upper limit in brightness so none will be found at their nominal brightness. The PI SNe that appear in this survey are all of 150~-~250~\Ms\ red supergiants (RSGs). These numbers can be partly understood from Figure~\ref{fig:lightcurve}. Even with the deeper exposure the unlensed CC SN is only visible for a short time in the $z$ band, and it is much less likely to be lensed because the FoV is so small in comparison to the WDF. On the other hand, PI SNe are visible for much longer times so the ADS can detect them in spite of its small FoV. Like CC SNe, they will not be lensed because only a fraction of the small FoV will be magnified.

\begin{figure}
\begin{center}
\includegraphics[width=80mm]{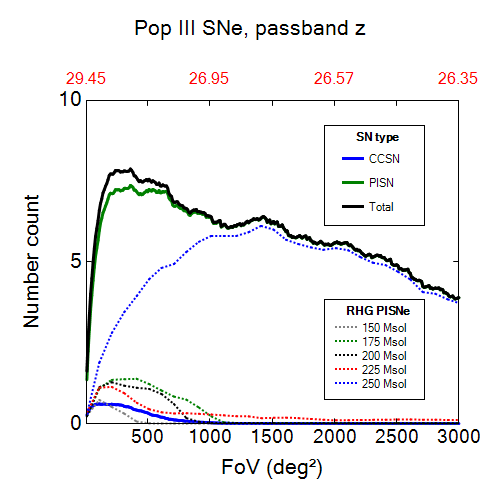}
\end{center}
\caption{
Pop III SN number counts in the $z$ band in the ADS as a function of FoV and sensitivity for a total exposure time of 640h. The lower $x$-axis is FoV in deg$^2$ and corresponding AB mag limits are noted on the upper axis in red. The CC SN number counts shown are for their upper limit in brightness.
}
\label{fig:ads_opt}
\end{figure}

640h is the minimum exposure required to find at least one Pop~III SN in a single 10~deg$^2$ field. In Figure~\ref{fig:ads_opt} we show cumulative Pop~III SN number counts over $z~\sim$~5~-~7 as a function of survey area and AB magnitude limit, again assuming a cadence of one year.  They peak at $\sim$~8 at a FoV of 250~deg$^2$ and m$_{\mathrm{AB}} = 27.6$. PI SNe dominate these counts, $\sim$~7 versus $\sim$~1 CC SN. We also show PI SN counts by progenitor mass, all of which are RSGs because explosions of blue supergiants (BSGs) are too faint to be seen. The 250~\Ms\ PI SNe will be most numerous ($\sim$~3) but $\sim$~4 of them will be 150~-~200~\Ms\ PI SNe. The number of 250~\Ms\ PI SNe increases with FoV, completely dominating the number count above 1000~deg$^2$.

\subsection{Sources of Uncertainty}
\label{sec:disc}

The primary source of uncertainty in our SN number counts is the rest frame SNRs, which are derived from cosmic SFRs and the assumed IMF. The global SFRs inferred from GRB observations by \citet{re12} vary by $\sim$ 20\% at $z =$ 5 to a factor of 2 - 3 at $z =$ 7. Since our SN number counts are directly proportional to the SFRs they vary by the same factor in this parameter. We adopted a Salpeter IMF for the Pop II/I stars in our models to obtain  conservative counts. If we instead use the more realistic Kroupa IMF we obtain counts that are $\sim$ 50\% higher. There is a large uncertainty in the IMF of Pop III stars but SN counts in the main survey are dominated by Pop I and II explosions so this uncertainty does not impact the total counts. The ADS is much more sensitive to uncertainties in the Pop III IMF. 

\subsubsection{SN Type}

Since at most 1\% of CC SNe are Type IIne and on average only 0.5 Pop~II would be found in the WDF, the actual fraction of CC SNe we take to be Type IIn has little impact on our counts predicted for this survey. This is not true of deep-drilling fields deeper than AB-magnitude $\gtrsim$~28 in which $>10$ Type IIn are are predicted to be found. Here, the uncertainty in ratio of Type IIne to all CC events becomes significant. Less than one PPI or RPI SN will be observed, so the number of $107.5-112.5$ \Ms\ stars we take to die as PPI events rather than RPI SNe would not alter our counts. We find that only RSG CC and PI SNe will be observed so  our predicted number counts are linearly dependent on the fraction of stars that die as RSGs, 75\% in our study. At the lower limit of luminosity no CC SN detections are predicted. At the upper limit, the number counts are all due to 25~\Ms\ RSG SNe, with the more energetic events dominating the total count. Thus, this count depends on the fractions of SNe assumed to be 0.6, 1.2 or 2.4 foe, which are equal in our study. Essentially all the PI SN detections are  250~\Ms\ RSG explosions.

\subsubsection{Detection Criteria}

In our detection criteria we flag every object whose flux has changed by 25\% or more as a SN. Changing this threshold has no effect on the predicted numbers of SNe in the WDF because it only tallies new events because the area on the sky being surveyed never changes.  Any new detection at the site of a previous one due to a change in its light curve is simply discarded, with no effect on the count. However, the ADS counts evolving sources as new SNe and is therefore sensitive to the threshold. But this effect is small. We find that changing the threshold from 25\% to 1\% only increases the number count by $\sim 0.01$. 

There is a dependence of number count on the cadence, and it can be used to set the actual cadence. Although the cadence of the main survey is already fixed at 27d on average, number counts in the ADS increase until the cadence reaches a few times 10 days and are more or less flat thereafter. Consequently, number counts in the ADS are not sensitive to cadence changes above 1 - 2 months.

\subsubsection{Scattering / Absorption}

We treat attenuation of flux due to Ly$\alpha$ scattering with the model of \citet{madau95} at $z < 6$ and with \citet{gp65} for $z > 6$. The errors in \citet{madau95} can be large but $\lambda_{\mathrm{Ly\alpha}}$ is only redshifted through the $i$ band and $\sim 35\%$ of the $z$ band at $z < 6$. The large uncertainties in \citet{madau95} therefore have little impact on our results because they only affect part of the $z$ band. High-redshift SNe might be obscured by dust at lower galactic altitudes. However, this will likely have little effect on the number of detections since this region of the sky is not covered by the WDF. There may also be extinction in the host galaxies. This extinction depends on the metallicity. Pop III galaxies are believed to be free of dust, so they would not obscure explosions \citep{ryd17a}. Dust in the host galaxies of Pop II SNe may dim them but the extinction is expected to be weak because they are being observed in rest-frame UV. At $z \sim$ 5 - 7 the host galaxies also likely have low metallicities and less dust.

\subsubsection{Lensing Model}

A source of potentially large uncertainty in our lensing models is the fraction of the sky with $1~<~\mu~<~ 2$ (Section~\ref{sec:mu}). The interpolation we use yields a rather severe lower limit because at $\mu = 1$ it returns a lensed area that is just a few percent of the sky. As a simple test of how robust our counts are with respect to this interpolation we have done it with a second-degree polynomial in the log$_{10}$ - log$_{10}$ plane as well. As constraints we use $f(\mu=1) = 1$ and $f(\mu=2) = f_2$, where f($\mu$) is the fraction of the sky magnified by at least $\mu$. The slope of the interpolation between $\mu=2$ and $\mu=3$ is set equal to the derivative of the polynomial at $\mu=2$. 

None of the counts in the WDF change because detections there always require magnifications greater than 2, as seen in Table~\ref{tab:mostluminous}. This is true even for CC SNe at their upper limit in brightness, which are not listed in Table~\ref{tab:mostluminous} but whose minimum magnifications can be obtained from those in the table by dividing them by ten. In the ADS the use of this interpolation adds on average only one lensed SN. Although this interpolation is arbitrary, it demonstrates the sensitivity of the counts to the magnification model at $1<\mu<2$.

Multiple images of SNe behind galaxies and galaxy clusters can confuse SN number counts. The appearance of such images can be separated by weeks because of time delays due to lensing, and because they appear in different regions of the sky they could be mistaken for distinct events. But unlensed explosions do not exhibit multiple images so this phenomenon will have no effect on their number counts. Time delays could boost detections of lensed explosions, but we neglect them because their numbers are expected to be small at low magnifications and because the number of counts that require high magnification are low, so even a relatively large boost would not change them much.

\subsection{Other Uncertainties}

Other factors can alter or reduce our number counts, such as choice of radiation transfer algorithm, other theoretical uncertainties in model spectra, and microlensing of lensed images. These factors can come into play especially in surveys with only nominal predictions of SN counts, such as the 10 deg$^2$ ADS in which only 1 - 2 explosions were expected to be found. Borderline number counts such as these are also subject to statistical effects that could reduce the actual yield of lensed Pop III SNe to zero. For example, the  $\sqrt{N}$ statistical error associated with predictions of 1 - 2 events yields 1 $\pm$ 1 or 2 $\pm$ 1.41 detections, which are consistent with zero.

\section{Conclusion}
\label{sec:conclusion}

We find that the LSST WDF survey will on average detect many Pop~II SNe from $z \sim$ 5 - 7, up to $\sim\popIIcczhigh{}$ CC SNe and $\sim \popIIpiz{}$ PI SNe, but no Pop III SNe. These are all lensed events because the WDF lacks the sensitivity required to directly detect SNe at these redshifts, even PI SNe. The absence of Pop~III SNe in the WDF is due to the much larger numbers of chemically enriched than pristine stars at this epoch. The large range in Pop~II CC SN count (0-\popIIcczhigh{}) is due to its highly nonlinear dependence on the threshold magnification required for detection. The factor of 10 by which CC SNe are likely to vary in luminosity leads to a factor of $\gtrsim$~100 in number count. Each Pop~III and Pop~II SN will be flagged on average twice due to variations in their light curves. We adopted uniform cadences for the WDF for simplicity.  While this introduces some error into our rates it is subsumed by the uncertainty in cosmic SFR at $z \sim$ 5 - 7, which is much higher.

A DD survey in a single 10~deg$^2$ FoV with two exposures one year apart must reach an AB-magnitude $> 26$ to observe Pop II~PI SNe at $z \sim$~5~-~7 and $>29$ to detect Pop~III PI SNe. For Pop~II CC SNe the depth required is AB mag 27 - 28. With a deeper exposure (640h) distributed over a 300~deg$^2$ FoV yielding m$_{\mathrm{AB}} = 27.6$, the ADS could discover $\sim 7$ Pop~III PI SNe and $\sim 1$ Pop~III CC SN.

After an event is flagged as a transient one can make an initial redshift cut from its Lyman break in one of the filters. If it is still bright enough after discovery by LSST it can then be studied in greater detail with {\em JWST} or one of the ELTs. A trigger on the degree of variation of the LC with cadence could also be set, given that high-$z$ SNe LCs will vary more slowly. But spectroscopic followup with the {\em JWST} NIRCam or ground based instruments would still be the best route to determining its redshift. As to whether or not a given event is lensed, one could either look for multiple images of the transient or compare its observed flux to that expected from the template explosion to which it is matched, given its source redshift.

Although we have shown that LSST can find Pop III SNe it is not currently possible to discriminate them from the much larger pool of transients LSST would discover at 5 $< z <$ 7 because even primordial SNe may exhibit metal lines at some point of their evolution due to  rotational or semi-convective mixing in the star prior to death. Even if they did not, line emission due to metals might be triggered if the ejecta expands through gas enriched by heavier elements, which would be likely at z $\sim$ 5 - 7. Consequently, although the likelihood that a given explosion is a Pop~III SN rises with redshift it becomes unambiguous only at $z \sim 20 - 25$, when there has not been sufficient time to pollute the cosmos with metals. For now, because the LSST $y$ and $z$ bands limit detections to $z \sim 5 - 7$, it could only be said statistically that some of them would be due to metal-free stars.

In the long term, observations by JWST and the ELTs at 2~-~5~$\mu$m will be required to find the first SNe in the universe. But wide-field surveys in the $H$ band by {\em Euclid} and the {\em Wide Field Infrared Survey Telescope} ({\em WFIRST}) could extend detections of lensed SNe to $z \sim$~15 in the interim and probe the properties of stars in the earliest galaxies. Calculations of lensed SN rates for these two missions at this era are now under development.

\section*{Acknowledgments}

We thank the anonymous referee, whose comments improved the quality of this paper. We also thank Daniel Holz and Matthias Bartelmann for their advice over the course of this project. CER was supported by the European Research Council under the European Community's Seventh Framework Programme (FP7/2007 - 2013) via the ERC Advanced Grant ``STARLIGHT: Formation of the First Stars" (project number 339177). DJW was supported by STFC New Applicant Grant ST/P000509/1. TC was funded by a University of Portsmouth Dennis Sciama Fellowship. Maturi and MC were partially supported by the Transregional Collaborative Research Centre TRR 33.




\bibliographystyle{mnras}
\bibliography{refs}



\bsp	
\label{lastpage}
\end{document}